\documentclass[12pt,a4paper]{article}
\usepackage{amsmath}
\usepackage{amsfonts}
\usepackage{amssymb}
\usepackage{amsthm}
\usepackage{physics}
\usepackage{jheppub}
\usepackage{bm}
\usepackage{graphicx}
\usepackage{url}
\usepackage{hyperref}
\usepackage{lmodern}
\usepackage{subfigure}
\usepackage[section]{placeins}
\usepackage[table,xcdraw]{xcolor}
\usepackage{float}
\usepackage{multirow}
\usepackage[utf8]{inputenc}
\usepackage[T1]{fontenc}
\usepackage{hyperref}
\hypersetup{
  colorlinks=true,
  citecolor=purple,
  linkcolor=blue,
  urlcolor=violet
 }

\newcommand{\bi}{\begin{itemize}}
\newcommand{\ei}{\end{itemize}}

\newcommand{\bea}{\begin{eqnarray}}
\newcommand{\eea}{\end{eqnarray}}
\newcommand{\be}{\begin{equation}}
\newcommand{\ee}{\end{equation}}
\newcommand{\ben}{\begin{eqnarray*}}
\newcommand{\een}{\end{eqnarray*}}
\newcommand{\bem}{\begin{pmatrix}}
\newcommand{\eem}{\end{pmatrix}}
\newcommand{\bl}{\begin{align}}
\newcommand{\el}{\end{align}}
\newcommand{\beg}{\begin{gather}}
\newcommand{\eeg}{\end{gather}}

\newcommand{\mi}{\mathrm{i}}
 
\newcommand{\cH}{\mathcal{H}}

\newcommand{\IH}{\mathbb{H}}

\newcommand{\TrH[1]}{ {\raise -.5em
                      \hbox{$\buildrel {\textstyle  {\rm Tr } }\over
{\scriptscriptstyle \cH _ {#1}}$}~}}
\newcommand{\res[1]}{{\raise -.5em 
\hbox{$\buildrel{\textstyle{\rm Res}}\over {\scriptscriptstyle {#1}}$}}}
\newcommand{\tends[1]}{{\raise -.5em 
\hbox{$\buildrel{\longrightarrow}\over {\scriptscriptstyle {#1}}$}}}

\def\dbend{\lower3.5pt\hbox{\manual\char127}}

\def\IL{\relax{\rm I\kern-.18em L}}
\def\IH{\relax{\rm I\kern-.18em H}}
\def\rlx{\relax\leavevmode}

\def\ZZ{\rlx\leavevmode\ifmmode\mathchoice{\hbox{\cmss Z\kern-.4em Z}}
 {\hbox{\cmss Z\kern-.4em Z}}{\lower.9pt\hbox{\cmsss Z\kern-.36em Z}}
 {\lower1.2pt\hbox{\cmsss Z\kern-.36em Z}}\else{\cmss Z\kern-.4em
 Z}\fi}

\title{\center{ \fontfamily{lmr}\selectfont Heterotic Strings \\ and \\ Quantum Entanglement}}

\preprint{}
\author{ \center \fontfamily{lmr}\selectfont  Atish  Dabholkar and} \author{\fontfamily{lmr}\selectfont  Upamanyu Moitra}
\affiliation{\begin{center}
The Abdus Salam International Centre for Theoretical Physics\\ Strada Costiera 11, Trieste 34151, Italy
\end{center}  }

\abstract{We construct $\mathbb{Z}_N$ orbifolds of the ten-dimensional heterotic string theories appropriate for implementing the stringy replica method for the calculation of quantum entanglement entropy.  A novel feature for the heterotic string is that the gauge symmetry must be broken by a Wilson line to ensure modular invariance.  We completely classify the patterns of symmetry breaking.  We show that the tachyonic contributions in all cases can be analytically continued, with a finite answer in the domain $0<N \leq 1$,  relevant for calculating entanglement entropy across the Rindler horizon.  We discuss the physical implications of our results.}

\makeatletter
\gdef\@fpheader{}
\makeatother
\begin{document}

\maketitle

\section{Introduction}\label{sec-intro}

In this article,  we extend the orbifold method \cite{Dabholkar:1994ai, Dabholkar:2001if,  Dabholkar:2022mxo} for computing the quantum entanglement entropy to heterotic string theory.  To this end,  we construct the $\mathbb{Z}_N$ orbifolds of the two-dimensional Rindler plane where $N$ is an odd positive integer.  One encounters a number of new features in this construction.  Unlike the orbifolds of Type-II theory \cite{Dabholkar:2023ows} and its compactifications \cite{Dabholkar:2023yqc},  and the orbifolds of the BTZ black hole in AdS$_3\times S^3 \times T^4$ \cite{Dabholkar:2023tzd} considered earlier,  the heterotic orbifolds are chiral.  As a result,  there is a potential for global gravitational anomalies on the world-sheet.  

As for heterotic Calabi-Yau compactifications,  modular invariant $\mathbb{Z}_N$ orbifolds are possible only if the gauge symmetry is broken in specific ways, essentially by embedding the spin connection in the gauge connection. Following standard orbifold methods \cite{Dixon:1985jw,  Dixon:1986jc} in the fermionic representation,  we give a complete classification of possible orbifolds both for the ${\rm Spin}(32)/\mathbb{Z}_2$ and the $E_8 \times E_8$ heterotic strings \cite{Gross:1984dd} for general $N$.

The main motivation for constructing these orbifolds is to implement a generalization of the replica method in string theory.  As explained in \cite{Dabholkar:2023ows},   the resulting entanglement entropy is  expected to be naturally UV-finite;  however,  there are potential IR divergences \cite{Witten:2018xfj, Dabholkar:2023ows,  Dabholkar:2023yqc, Dabholkar:2023tzd}  because of the presence of tachyons in the orbifolded theory.   Following the methods first described in \cite{Dabholkar:2023ows},  we analyze the tachyonic terms in all these cases. The analysis in each case is quite distinct but, remarkably, in all cases, a suitable analytic continuation  in $N$ exists for the tachyonic piece which is finite  in the physical region $0<N \leq 1$. 

Since there are many different ways of breaking the gauge symmetry, there are many different partition functions and their possible analytic continuations.  It  is  thus not  obvious that there will be a unique answer for the entanglement entropy.  One possibility is that the entanglement entropy obtained by a derivative of the partition function at $N=1$ is nevertheless unique. Another  physically reasonable possibility is that these different answers correspond to different superselection sectors. It is well known that in the computation of entanglement entropy in a gauge theory,  there are different superselection sectors  \cite{Buividovich:2008gq,  Donnelly:2011hn,  Casini:2013rba,  Ghosh:2015iwa,  Aoki:2015bsa,  Donnelly:2015hxa,  Blommaert:2018oue, Moitra:2018lxn,  Moitra:2022ooc,  Ball:2024hqe} corresponding to different values of the normal component of the electric field at the horizon because of the Gauss law constraint (see also  \cite{Kabat:1995eq}). It is thus possible that string theory dictates a specific choice of these superselection sectors. We leave the investigation of these issues for the future.

\section{The Heterotic Orbifolds}\label{sec-hetorb}

We construct $\mathbb{Z}_N$  orbifolds of the  ten-dimensional heterotic string in the fermionic formulation and classify the possible choices for breaking the gauge symmetry. The target space is  $(\mathbb{R}^2 / \mathbb{Z}_N) \times \mathbb{R}^8$.  An analytic continuation in $N$, if it exists, would then allow one to formally obtain the replica partition function branched at the Rindler horizon at the origin of $\mathbb{R}^2$.

The heterotic string  combines the right-moving sector of the 10-dimensional superstring with the left-moving sector of the 26-dimensional bosonic string.  In the light-cone gauge,  the right-moving sector consists of 8 free bosons and 8 free fermions in the Ramond-Neveu-Schwarz (RNS) formalism.  In the fermionic construction,  the left-moving sector consists of 8 free  bosons and 32 real fermions with suitable boundary conditions \cite{Polchinski:1998rr}.  

The $\mathbb{Z}_N$ group acts on the Rindler plane $\mathbb{R}^2$ spanned by two bosonic coordinates, $N$ being an odd positive integer.  There is an equivalent action on a pair of right-moving fermions because of the superconformal symmetry on the world-sheet.  We package the two bosonic Rindler coordinates $X^1$ and $X^2$ in a single complex coordinate $X = X^1 + \mi X^2$ and its complex conjugate.   Let $\psi$ be the corresponding right-moving complex fermion. There are $N$ twisted sectors of the orbifold.  In the twisted sector labeled by $k$, the orbifold generator $g$ acts on $X$  with the twist $2k/N$
\begin{equation}
g: \quad 
\bqty{X \atop \psi}
 \to e^{ \frac{4\pi \mi k}{N} } \bqty{X \atop \psi}.  \label{xpsit}
\end{equation}
We will see that modular invariance requires an appropriate action on the left-moving internal fermions as well.

As in \cite{Dabholkar:2023ows,  Dabholkar:2023yqc,  Dabholkar:2023tzd},  we are interested in the genus-one string partition function
\begin{equation}
Z^{(1)} (N) = A_H \int\limits_{\mathcal{D}} \frac{\dd[2]{\tau}}{\tau_2^2} \mathcal{F} (\tau, N),
\end{equation}
where $\mathcal{D}$ is the standard fundamental domain (``keyhole region'') of the torus modular parameter $\tau \equiv \tau_1 + \mi \tau_2$ and $A_H$ is the regulated area of the Rindler horizon. 

The full partition function for both the heterotic string theories involves a double sum with $k$ labeling the twisted sectors and $\ell$ denoting the action of twining by $g^\ell$ (enforcing a projection to $\mathbb{Z}_N$-invariant states),
\begin{equation}
\mathcal{F} (\tau) = \frac{1}{N}\frac{1}{\tau_2^3} \sum_{k, \ell \in \mathbb{Z}_N }  \mathcal{F}_{k\ell; R} (\tau) \overline{\mathcal{F}_{k\ell; L} (\tau)}. \label{ztzrzl2}
\end{equation}

The right-moving sector has the same partition function in every case,
\begin{equation}
\begin{aligned}
 \mathcal{F}_{k\ell; R} (\tau) &=  \frac{ \sum_{a =1}^4 (-1)^{a+1} \vartheta_a \qty( \frac{2k\tau + 2\ell}{N} \Big| \tau ) \vartheta_a^3 \qty(0 | \tau )}{2\eta^{9} (\tau) \vartheta_1\qty( \frac{2k\tau + 2\ell}{N} \big| \tau )}  \\
 &= \frac{\vartheta_1^4 \qty( \frac{k\tau + \ell}{N} \big| \tau )}{\eta^{9} (\tau) \vartheta_1\qty( \frac{2k\tau + 2\ell}{N} \big| \tau )}.
 \end{aligned} \label{zklr}
\end{equation}
The first line corresponds to the partition function for the right-movers in the RNS formalism where  we have  performed the Gliozzi-Scherk-Olive (GSO) projection \cite{Gliozzi:1976qd} summing over spin structures. The second line corresponds to the partition function in the Green-Schwarz formalism, equivalent to the first line by a quartic Riemann identity (see,  e.g.,  eq.  (5.8) of \cite{Dabholkar:2023yqc}).  The case $N = 1$ corresponds to un-orbifolded ten-dimensional theory  and the partition function vanishes,   $\mathcal{F}_{00; R} (\tau) = 0$,  on account of supersymmetry.

Let us now discuss the left-movers, first for the $\mathrm{Spin}(32)/\mathbb{Z}_2$ string.  As mentioned previously,  in addition to 8 left-moving bosons,  we also have 32 real fermions which we repackage into 16 complex fermions $\lambda^A$  $(A = 1, 2, \cdots, 16)$.   In the ordinary heterotic string,  these fermions obey either Ramond (R) or Neveu-Schwarz (NS) boundary conditions. 

For the $\mathbb{Z}_N$ orbifold, one can have more general  order $N$ twists for  internal fermions:
\begin{equation}
\lambda^A ( \sigma^1 + 2\pi ,  \sigma^2) = \pm \exp( 2 \pi \mi \frac{k_A}{N}  )   \lambda^A ( \sigma^1,  \sigma^2),  \label{lambtr}
\end{equation}
where the upper (lower) sign holds for the R (NS) sector. For now, $\{ k_A \}$ are arbitrary integers to be determined  by requiring modular invariance.\footnote{Every $k_A$ simultaneously being half-odd integral is also an allowed possibility --- which is also order $N$ as long as both R and NS sectors are included.}

 One must perform the GSO projection in each sector  to get the modular invariant partition function. The general form of the partition function for the $1$ complex and $6$ real spacetime bosons and the $16$ complex internal fermions is given by \begin{equation}
\mathcal{F}_{k\ell;  L} (\tau) = \frac{\sum_{a = 1}^4 \prod_{A = 1}^{16} \vartheta_a \qty( \frac{k_A \tau + \ell_A }{N} \big| \tau ) }{2\eta^{21} (\tau) \vartheta_1\qty( \frac{2k\tau + 2\ell}{N} \big| \tau ) }.  \label{fklspin}
\end{equation}
where $\{ k_A \}$ are the twists and $\{\ell_A \}$ are the twines for the internal fermions.
The functions $\vartheta_{1,2}$ are associated with the R-sector Hilbert space and $\vartheta_{3,4}$  with the NS-sector. 

At first glance,   there  appears to be a large number of possibilities involving the set of twistings $\{k_A \}$.  As expected,   modular invariance \cite{Dixon:1986jc,  Vafa:1986wx} severely constrains the allowed possibilities as we now discuss. 
We look at the ground state energy of the left-movers in the Ramond sector and right-movers specified by the twists $\{ k_A \}$ and $k$.  On the right,  for $k < N/2$,  the complex boson contributes an energy $ - \frac{1}{12} + \frac{1}{2} \frac{2k}{N} \qty( 1 - \frac{2k}{N}  )$ and each complex fermion contributes  $ +\frac{1}{12} - \frac{1}{2} \frac{k}{N} \qty(1 - \frac{k}{N} )$.  Noting that there are 6 additional neutral real bosons each with energy $-\frac{1}{24}$ , we have the total energy
\begin{equation}
E_R =  - \frac{k}{N},
\end{equation}
as for the Type-II orbifolds \cite{Dabholkar:2023ows}.

In the Ramond sector of the left-movers, the bosonic contribution is the same but $\lambda_A$ contributes $+ \frac{1}{12} - \frac{1}{2} \frac{k_A}{N} \qty( 1 - \frac{k_A}{N}  )$,  giving the total energy
\begin{equation}
E_L =  1 + \frac{k}{N} - \frac{2k^2}{N^2}  - \frac{1}{2} \sum_{A=1}^{16} \frac{k_A}{N} \qty( 1- \frac{k_A}{N}  ) .
\end{equation}
Since the oscillators have $1/N$ moding, level-matching requires that  $N( E_L - E_R)$ is an integer. Therefore,
 \begin{equation}
2k - \frac{1}{2}\sum_{A =1}^{16} k_A  +  \frac{1}{2N}\sum_{A = 1}^{16} k_A ^2 - \frac{2k^2}{N}  = m,
\end{equation}
where $m$ is an integer.
The orbifold group must have order $N$ also acting on the spinors of $\mathrm{Spin} (32) / \mathbb{Z}_2$ which requires that $\sum_{A} k_A$ be even 
(for odd $N$, however, this condition can always be met by an appropriate choice of basis).  Thus,  $\sum_A k_A^2$ is even as well and in the end, one finds the condition \cite{Dixon:1986jc,  Vafa:1986wx} 
\begin{equation}
\sum_{A=1}^{16} k_A^2  - (2k)^2 \equiv  0 \, (\mathrm{mod} \, N ) . \label{condtwi}
\end{equation}

First note from \eqref{lambtr} that $k_A \to k_A + N$ is an exact symmetry of the problem.  
Using this symmetry,  we can choose a value of $k_A$ so that $|k_A| < N /2$.  All the $k_A$'s being half-odd integral meeting the above constraints are also allowed.  The properties of the Jacobi $\vartheta$-functions  also necessitate that if any $k_A$ is half an odd integer,  then so is every other $k_A$ for modular invariance.  Furthermore, given that we are interested in the patterns of gauge symmetry breaking that hold true for any $N$, we can impose a stronger condition  replacing the congruence relation \eqref{condtwi} with an equation
\begin{equation}
\sum_{A=1}^{16} k_A^2  - (2k)^2 = 0 . \label{conseq}
\end{equation}
For a given $N = N_1$,  it might be possible to find special solutions that satisfy \eqref{condtwi} but not \eqref{conseq}, even after appropriate shifts.   However,  these special solutions that are valid only modulo  $N = N_1$ would not be allowed for another generic $N = N_2$ under the $\mathrm{mod}\, N$ condition.  Thus,  with a view towards possible analytic continuations,  we settle on the equation \eqref{conseq} which of course remains true for any $N$.

For the  $\mathrm{Spin}(32)/\mathbb{Z}_2$ heterotic string,  there are only three distinct possibilities meeting \eqref{conseq}:
\begin{enumerate}
\item $\Bqty{{k}_{A}}= k \qty\Big{ {\frac{1}{2}}^{16} }$,  \quad $\{\ell_{A}\} = \ell \qty\Big{ {\frac{1}{2}}^{16} }$\, ;
\item $\{ k_{A}  \}= k \qty\Big{ 1^4,  0^{12} }$,  \quad $\{\ell_{A}\} = \ell \qty\Big{ 1^4,  0^{12} }$;
\item $\{k _{A} \}= k \qty\Big{ 2,  0^{15} }$,  \quad $\{\ell_{A}\}  = \ell \qty\Big{ 2,  0^{15} }$,
\end{enumerate}
where we see that the twines $\{\ell_A\}$ must be proportional to the same vectors as the twists $\{k_A\}$ to ensure modular invariance.  These vectors determine the pattern of symmetry breaking and their normalization depends on  $k$ and $\ell$. As a result, for a given symmetry-breaking pattern,  in the final answer for the partition function in \eqref{fklspin} and in
\eqref{ztzrzl2},  the sectors are labeled only by $k$ and $\ell$ without any additional labels that depend on $\{ k_A ,  \ell_A \}$. 

It is not a surprise that all the vectors above belong to the weight lattice of $\mathrm{Spin}(32) / \mathbb{Z}_2$. 
The first vector belongs to the spinor conjugacy class and the two latter belong to the adjoint conjugacy class.  One can incorporate changes in the sign (for the spinor class,  the number of sign changes must be even) and permutations of the components which yield the same partition function.

The analysis for the $E_8 \times E_8$ string is similar and straightforward.  In this case, the 16 complex internal left-moving fermions are split into two sets of 8 each with $\lambda^A$,  $\lambda^B$ $(A, B = 1,  2,  \cdots , 8)$. The GSO projection is implemented independently for each set.  As in \eqref{lambtr},  let us consider the twists $\{k_A/N \}$ and $\{ k'_B / N \}$ for the two sets of fermions. The resulting partition function is 
\begin{equation}
\mathcal{F}_{k\ell;  L} (\tau) = \frac{\sum_{a, b = 1}^4 \prod_{A = 1}^{8}  \vartheta_a \qty( \frac{k_A \tau + \ell_A }{N} \big| \tau ) \prod_{B = 1}^{8} \vartheta_b \qty( \frac{k'_B \tau + \ell'_B }{N} \big| \tau  ) }{4\eta^{21} (\tau) \vartheta_1\qty( \frac{2k\tau + 2\ell}{N} \big| \tau) }.  \label{fkle8}
\end{equation}
We will call the different left-moving sectors as R-R,  R-NS,  NS-R and NS-NS with the same correspondence with the $\vartheta$-functions as before. 

With similar arguments such as those used previously, we can narrow down the possible choices of the vectors $\{ k_A \} $ and $\{k'_B \}$ to five distinct cases:
\begin{enumerate}
\item $\{ k_A \}  = k \qty\Big{ {\frac{1}{2}}^{8} }, \quad \{ k'_B \}  = k \qty\Big{ {\frac{1}{2}}^{8} }$;
\item $\{ k_A \}  = k \qty\Big{ 1,  1,  0^{6} }, \quad \{ k'_B \}  = k \qty\Big{ {\frac{1}{2}}^{8} }$;
\item $\{ k_A \}  = k \qty\Big{ 1,  1,  0^{6} }, \quad \{ k'_B \}  = k \qty\Big{ 1,  1,  0^{6} }$;
\item $\{ k_A \}  = k \qty\Big{ 1^4,  0^{4} }, \quad \{ k'_B \}  =  \qty\Big{ 0^{8} }$;
\item $\{ k_A \}  = k \qty\Big{ 2,  0^{7} }, \quad \{ k'_B \}  =  \qty\Big{ 0^{8} }$.
\end{enumerate}
As before, appropriate sign changes and permutations lead to the same partition functions,  as does an interchange of the two vectors $\{ k_A \} $ and $\{k'_B \}$.  The vectors $\{\ell_A \}$ and $\{\ell'_B \}$ can be obtained by replacing $k$ with $\ell$ in the relations above.

Before we end this section,  let us remark on the $N = 1$ case, for which $k = 0 = \ell$. In this case,  
the denominator of the $\mathrm{Spin}(32) / \mathbb{Z}_2$ partition function \eqref{fklspin} becomes zero (corresponding to the bosonic zero mode giving the volume of the $\mathbb{R}^2$).  Regularizing the infinite volume factor and concentrating only on the oscillators one obtains the partition function of the ten-dimensional un-orbifolded theory as expected.  
In particular,  the partition function for the left-moving fermions is
\begin{equation}
\mathcal{F}^{ \mathrm{Spin(32)/\mathbb{Z}_2} }_{L,  \,  \mathrm{fermionic}} (\tau )  = \frac{1}{2 \eta^{16} (\tau)} \pqty{ \vartheta_2^{16} (\tau ) +  \vartheta_3^{16} (\tau) +  \vartheta_{4}^{16} (\tau )  } .   \label{zltauspin}
\end{equation}
From \eqref{fkle8} we analogously find for the $E_8 \times E_8$ case that
\begin{equation}
\mathcal{F}^{ E_8 \times E_8 }_{L,  \,  \mathrm{fermionic}}  (\tau )  = \frac{1}{4 \eta^{16} (\tau)} \pqty{ \vartheta_2^{8} (\tau ) +  \vartheta_3^{8} (\tau) +  \vartheta_{4}^{8} (\tau )  }^2. \label{zltaue8}
\end{equation}
It is easy to check that the partition functions for the two different string theories are the same by appealing to the aforementioned quartic Riemann identity,
\begin{equation}
\mathcal{F}^{ E_8 \times E_8 }_{L,  \,  \mathrm{fermionic}}  (\tau )  = \mathcal{F}^{ \mathrm{Spin(32)/\mathbb{Z}_2} }_{L,  \,  \mathrm{fermionic}}  (\tau ).\label{eqzse}
\end{equation}
Of course, this feature is not expected to persist for the orbifolds once the gauge symmetry is broken.

\section{Analytic Continuation of Tachyons}

We now isolate the tachyonic parts of $\mathcal{F} (\tau, N)$, growing exponentially with large $\tau_2$ or equivalently,  for small values of $q\equiv \exp(2\pi\mi\tau)$ with a thorough analysis of the string spectrum.  As in the case of the ten-dimensional Type-II superstring \cite{Dabholkar:2023ows},  by grouping some non-tachyonic states with the tachyons in a suitable manner, it is possible to split the integrand as 
\begin{equation}
\mathcal{F} (\tau, N) = \widetilde{\mathcal{F}}^T_0 (\tau_2,  N) + \widetilde{\mathcal{F}}^R (\tau,  N),
\end{equation} 
where the function $\widetilde{\mathcal{F}}^T_0 (\tau_2,  N)$ is \emph{analytic} in $N$ and includes the contribution of all the tachyons in the spectrum. Moreover, as in all the other cases analyzed earlier, remarkably, the resulting analytic function  $\widetilde{\mathcal{F}}^T_0$  has the feature that it is non-tachyonic for $0<N\leq 1$ and yields a \emph{finite} integral in this region. After subtracting  the tachyons,  the remaining part of the integral involving $\widetilde{\mathcal{F}}^R (\tau,  N)$ is manifestly finite for odd $N$ and can be evaluated numerically to find a suitable extrapolation.
 Thus,  in the physical regime, $0<N \leq 1$,  the total modular integral can be argued to be finite \cite{Dabholkar:2023ows}.

It is noteworthy that the tachyonic content for all heterotic cases above can be expressed in terms of just two such functions:
\begin{align}
 \widetilde{\mathcal{F}}^T_{0,\alpha} (\tau_2 , N) &= - \frac{2}{\tau_2^3} \sum_{n = 0}^\infty e^{-4\pi \tau_2 n} \frac{1 - e^{  \frac{N-1}{N} (2n +1)  2\pi \tau_2 } }{1 - e^{- \frac{1}{N} (2n+1) 4\pi \tau_2} } ,  \label{ftoa}\\
   \widetilde{\mathcal{F}}^T_{0, \beta} (\tau_2 , N)  &=  \widetilde{\mathcal{F}}^T_{0,\alpha} (\tau_2 , N) + \frac{2}{\tau_2^3}  \frac{1-e^{2 \pi \tau_2 \frac{ (N-1)}{N}}}{1 - e^{-4 \pi  \tau_2\frac{1}{N}}} . \label{ftob}
\end{align}
The first function $ \widetilde{\mathcal{F}}^T_{0,\alpha} (\tau_2 , N)$ is exactly the same that arose after resumming the tachyonic contribution with additional states in the Type-II theory in \cite{Dabholkar:2023ows}.

\subsection[Tachyons in   $\mathrm{Spin} (32) / \mathbb{Z}_2$ Orbifolds]{\boldmath Tachyons in   $\mathrm{Spin} (32) / \mathbb{Z}_2$ Orbifolds}

Let us now examine the three distinct choices of gauge symmetry breaking one by one.   For the tachyonic analysis,  we will always consider the half-range $1 \leq k \leq (N-1)/2$ --- the other half-range  has an identical spectrum.

\subsubsection*{\boldmath $\{ k_{A} \} = k \qty\Big{ {\frac{1}{2}}^{16} }$} 

Let us present some details of the analysis in this case.   The gauge symmetry is broken down to $\mathrm{U} (16)$.  It is clear that the leading order term in a small-$q$ expansion comes from the NS sector.  The ground state energies on the right and left are $-k/N$ and $(-1 + k/N)$ respectively.  These states are neither level-matched nor $\mathbb{Z}_N$-invariant. Thus, they do not jointly contribute to the physical spectrum.  The fractionally moded bosonic oscillators with the mode number $(1-2k/N)$ will raise the energy of the ground state on both sides. Acting with such an oscillator on the left ground state gives an excited state of energy $-k/N$, which is level-matched with the right-side ground state.  

We can create further level-matched and $\mathbb{Z}_N$-invariant sub-leading tachyons by acting with the fractional bosonic oscillators on each side.
As was also explained in \cite{Dabholkar:2023ows,  Dabholkar:2023yqc, Dabholkar:2023tzd} the contribution in which the bosonic oscillators are added infinitely many times on the tachyonic state of energy $-k/N$ can be written as
\begin{equation}
 \widetilde{\mathcal{F}}^T_{0,\alpha} (\tau_2 , N) = \frac{2}{\tau_2^3} \sum_{k=1}^{ \frac{N-1}{2} } e^{ \frac{2k}{N} 2\pi \tau_2 } \sum_{n=0}^\infty e^{ -2 n \qty(1 - \frac{2k}{N} ) 2\pi \tau_2  }.  \label{ftoa2}
\end{equation}
The factor of 2 arises because of the $k\leftrightarrow N-k$ symmetry of the spectrum.   One can convince oneself that the entire spectrum of physical tachyons associated with the left-moving NS sector is included in this sum. For a fixed $\tau_2$, interchanging the two sums and performing the finite geometric sum in $k$,  we recover the analytic expression given in \eqref{ftoa}.

Let us now consider the left-moving Ramond sector.  The ground state energy is $(1- 3k/N)$ which is tachyonic for $k > N/3$.  Again,  this is not level-matched with the ground state on the right.  However,  if we act with the fractional bosonic oscillator on the right,  we find an excited state of energy $(1-3k/N)$ which is level-matched with the R ground state on the left. We can form a similar tower with fractionally moded oscillators on both sides.  The net contribution can be written as
\begin{equation}
 \widetilde{\mathcal{F}}^T_{0,\beta} (\tau_2 , N) = \frac{2}{\tau_2^3} \sum_{k=1}^{ \frac{N-1}{2} } e^{ - 2\qty(1- \frac{3k}{N} )2\pi \tau_2 } \sum_{n=0}^\infty e^{ -2 n \qty(1 - \frac{2k}{N} ) 2\pi \tau_2  }.  \label{ftob2}
\end{equation}
This contribution is analytically continued to the form given in \eqref{ftob}. 

In summary, the entire tachyonic contribution in this case can be written simply as
\begin{equation}
\widetilde{\mathcal{F}}^T_0 (\tau_2,  N) =  \widetilde{\mathcal{F}}^T_{0,\alpha} (\tau_2 , N) +  \widetilde{\mathcal{F}}^T_{0,\beta} (\tau_2 , N).
\end{equation}

\subsubsection*{\boldmath $\{ k_{A} \} = k \qty\Big{ 1^4,  0^{12} }$} 

The unbroken gauge group is $\mathrm{U} (4) \times \mathrm{SO} (24)$.   The R-sector ground state on the left is non-tachyonic.  Therefore,  the R-sector does not contribute to the tachyon spectrum.  As before,  the NS sector ground state is tachyonic and similar operations as before can be performed to yield $\widetilde{\mathcal{F}}^T_{0,\alpha}$.  Interestingly,  in contrast with the previous cases,  the fermionic oscillators associated with the numerator of \eqref{fklspin} make a contribution to the tachyonic spectrum: there are 6 possible ways to raise the left energy to $-k/N$, which is level-matched with the right with the net contribution $6\widetilde{\mathcal{F}}^T_{0,\alpha}$.  Finally, the fermionic oscillators give one excited state of energy $(1-3k/N)$,  which leads to one contribution of $\widetilde{\mathcal{F}}^T_{0,\beta}$.  Altogether,  we have
\begin{equation}
\widetilde{\mathcal{F}}^T_0 (\tau_2,  N) =  7\widetilde{\mathcal{F}}^T_{0,\alpha} (\tau_2 , N) +  \widetilde{\mathcal{F}}^T_{0,\beta} (\tau_2 , N).
\end{equation}

\subsubsection*{\boldmath $\{ k_{A} \} = k \qty\Big{ 2,   0^{15} }$} 
The unbroken gauge group is $\mathrm{U} (1) \times \mathrm{SO} (30)$.   The R-sector is non-tachyonic as in the previous case.  From the NS-sector,  taking into account the fermionic contributions,  we obtain the net tachyonic content
\begin{equation}
\widetilde{\mathcal{F}}^T_0 (\tau_2,  N) =  31\widetilde{\mathcal{F}}^T_{0,\alpha} (\tau_2 , N) +  \widetilde{\mathcal{F}}^T_{0,\beta} (\tau_2 , N).
\end{equation}

\subsection[Tachyons in  $E_8 \times E_8$ Orbifolds]{\boldmath Tachyons in  $E_8 \times E_8$ Orbifolds}

\subsubsection*{\boldmath $\{ k_A \}  = k \qty\Big{ {\frac{1}{2}}^{8} }, \quad  \vec{k}'_B = k \qty\Big{ {\frac{1}{2}}^{8} }$} 

The unbroken gauge group is $\mathrm{U} (8) \times \mathrm{U} (8)$. The R-R sector and NS-NS sectors have spectra similar with the R and NS sectors (respectively) of the $\mathrm{U} (16)$ case discussed previously.  Each of the R-NS and NS-R sector has the ground state energy $-k/N$,  which is level-matched with the right ground state and contributes to the same tachyonic tower involving $\widetilde{\mathcal{F}}^T_{0,\alpha}$.   The total tachyonic contribution is therefore
\begin{equation}
\widetilde{\mathcal{F}}^T_0 (\tau_2,  N) =  3\widetilde{\mathcal{F}}^T_{0,\alpha} (\tau_2 , N) +  \widetilde{\mathcal{F}}^T_{0,\beta} (\tau_2 , N).
\end{equation}

\subsubsection*{\boldmath $\{ k_A \}  = k \qty\Big{ 1, 1, 0^6 }, \quad  \vec{k}'_B = k \qty\Big{ {\frac{1}{2}}^{8} }$} 

The gauge symmetry is broken to $\mathrm{U} (2) \times \mathrm{SO} (12) \times \mathrm{U} (8)$.  The R-R  and R-NS sectors are both free of tachyons.  The NS-R sector contributes to two tachyonic towers involving $\widetilde{\mathcal{F}}^T_{0,\alpha}$ and $\widetilde{\mathcal{F}}^T_{0,\beta}$.  The NS-NS sector contributes to two towers of tachyons,  both involving $\widetilde{\mathcal{F}}^T_{0,\alpha}$.   The net contribution is
\begin{equation}
\widetilde{\mathcal{F}}^T_0 (\tau_2,  N) =  3\widetilde{\mathcal{F}}^T_{0,\alpha} (\tau_2 , N) +  \widetilde{\mathcal{F}}^T_{0,\beta} (\tau_2 , N).
\end{equation}

\subsubsection*{\boldmath $\{ k_A \}  = k \qty\Big{ 1, 1, 0^6 }, \quad  \vec{k}'_B =  k \qty\Big{ 1, 1, 0^6 } $} 

The unbroken gauge group is $\mathrm{U} (2) \times \mathrm{SO} (12) \times \mathrm{U} (2) \times \mathrm{SO} (12)$ in this case.  The R-R,  NS-R and R-NS sectors are all free of tachyons.  The only tachyons appear in the NS-NS sector. Taking into account the fermionic oscillators,  the net contribution is given by
\begin{equation}
\widetilde{\mathcal{F}}^T_0 (\tau_2,  N) =  3\widetilde{\mathcal{F}}^T_{0,\alpha} (\tau_2 , N) +  \widetilde{\mathcal{F}}^T_{0,\beta} (\tau_2 , N),
\end{equation}
which is the same as in the previous two cases.

\subsubsection*{\boldmath $\{ k_A \}  = k \qty\Big{ 1^4, 0^4 }, \quad  \vec{k}'_B =  k \qty\Big{0^8 } $} 

The unbroken gauge group is $\mathrm{U} (4) \times \mathrm{SO} (8) \times E_8$.  The R-R and NS-R sectors possess no tachyons. The R-NS and NS-NS sectors contribute to 8 and 7 tachyonic towers corresponding to $\widetilde{\mathcal{F}}^T_{0,\alpha} (\tau_2 , N)$ respectively. The NS-NS sector gives rise to another tower involving $\widetilde{\mathcal{F}}^T_{0,\beta} (\tau_2 , N)$. Thus,
\begin{equation}
\widetilde{\mathcal{F}}^T_0 (\tau_2,  N) =  15\widetilde{\mathcal{F}}^T_{0,\alpha} (\tau_2 , N) +  \widetilde{\mathcal{F}}^T_{0,\beta} (\tau_2 , N).
\end{equation}

\subsubsection*{\boldmath $\{ k_A \}  = k \qty\Big{ 2, 0^7 }, \quad  \vec{k}'_B =  k \qty\Big{0^8 } $} 

In this final case,  the gauge symmetry is broken to $\mathrm{U} (1) \times \mathrm{SO} (14) \times E_8$.   
As in one of the previous examples,  the R-R,  R-NS, NS-R sectors are tachyon-free.  A careful consideration of the NS-NS sector gives us the complete tachyonic contribution of the theory,
\begin{equation}
\widetilde{\mathcal{F}}^T_0 (\tau_2,  N) =  15\widetilde{\mathcal{F}}^T_{0,\alpha} (\tau_2 , N) +  \widetilde{\mathcal{F}}^T_{0,\beta} (\tau_2 , N).
\end{equation}

\begin{table}[]
\resizebox{\textwidth}{!}
{\begin{tabular}{|c|c|c|c|}
\hline
\rowcolor[HTML]{EFEFEF} 
\begin{tabular}[c]{@{}c@{}}Parent Theory\end{tabular} & \begin{tabular}[c]{@{}c@{}}Twist  Vector(s)\end{tabular} & \begin{tabular}[c]{@{}c@{}}Gauge Group \end{tabular} & \begin{tabular}[c]{@{}c@{}}Tachyonic Term $\widetilde{\mathcal{F}}^T_0 (\tau_2,  N) $ \end{tabular} \\ \hline \hline
                                                       & $\{ k_A \}  = k \qty\Big{ {\frac{1}{2}}^{16} }$                                                                 & $\mathrm{U} (16)$                                                      & $\widetilde{\mathcal{F}}^T_{0,\alpha} (\tau_2 , N) +  \widetilde{\mathcal{F}}^T_{0,\beta} (\tau_2 , N)$                                                     \\ \cline{2-4} 
                                                       & $\{ k_A \}  = k \qty\Big{ 1^4,  0^{12} }$                                                                 & $\mathrm{U} (4) \times \mathrm{SO} (24)$                                                      & $ 7\widetilde{\mathcal{F}}^T_{0,\alpha} (\tau_2 , N) +  \widetilde{\mathcal{F}}^T_{0,\beta} (\tau_2 , N)$                                                     \\ \cline{2-4} 
\multirow{-3}{*}{$\mathrm{Spin}(32)/\mathbb{Z}_2$}                                  & $\{ k_A \}  = k \qty\Big{ 2,  0^{15} }$                                                                 & $\mathrm{U} (1) \times \mathrm{SO} (30)$                                                     & $31\widetilde{\mathcal{F}}^T_{0,\alpha} (\tau_2 , N) +  \widetilde{\mathcal{F}}^T_{0,\beta} (\tau_2 , N)                                                    $ \\ \hline
                                                       & \begin{tabular}[c]{@{}c@{}} $\{ k_A \}  = k \qty\Big{ {\frac{1}{2}}^{8} }, \quad \{ k'_B \}  = k \qty\Big{ {\frac{1}{2}}^{8} }$ \end{tabular}                     & $\mathrm{U} (8) \times \mathrm{U} (8)$                                                      & $ 3\widetilde{\mathcal{F}}^T_{0,\alpha} (\tau_2 , N) +  \widetilde{\mathcal{F}}^T_{0,\beta} (\tau_2 , N)$                                                     \\ \cline{2-4} 
                                                       & \begin{tabular}[c]{@{}c@{}}$\{ k_A \}  = k \qty\Big{ 1,  1,  0^{6} }, \quad \{ k'_B \}  = k \qty\Big{ {\frac{1}{2}}^{8} }$\end{tabular}                     & $\mathrm{U} (2) \times \mathrm{SO} (12) \times \mathrm{U} (8)$.                                                     &  $ 3\widetilde{\mathcal{F}}^T_{0,\alpha} (\tau_2 , N) +  \widetilde{\mathcal{F}}^T_{0,\beta} (\tau_2 , N)$                                                    \\ \cline{2-4} 
                                                       & \begin{tabular}[c]{@{}c@{}} $\{ k_A \}  = k \qty\Big{ 1,  1,  0^{6} }, \quad \{ k'_B \}  = k \qty\Big{ 1,  1,  0^{6} }$ \end{tabular}                     & $\mathrm{U} (2) \times \mathrm{SO} (12) \times \mathrm{U} (2) \times \mathrm{SO} (12)$                                                     & $ 3\widetilde{\mathcal{F}}^T_{0,\alpha} (\tau_2 , N) +  \widetilde{\mathcal{F}}^T_{0,\beta} (\tau_2 , N)$                                                       \\ \cline{2-4} 
                                                       & \begin{tabular}[c]{@{}c@{}}$\{ k_A \}  = k \qty\Big{ 1^4,  0^{4} }, \quad \{ k'_B \}  =  \qty\Big{ 0^{8} }$\end{tabular}                     & $\mathrm{U} (4) \times \mathrm{SO} (8) \times E_8$                                                      & $ 15\widetilde{\mathcal{F}}^T_{0,\alpha} (\tau_2 , N) +  \widetilde{\mathcal{F}}^T_{0,\beta} (\tau_2 , N)$                                                       \\ \cline{2-4} 
\multirow{-5}{*}{$E_8\times E_8$}                                  & \begin{tabular}[c]{@{}c@{}} $\{ k_A \}  = k \qty\Big{ 2,  0^{7} }, \quad \{ k'_B \}  =  \qty\Big{ 0^{8} }$\end{tabular}                     & $\mathrm{U} (1) \times \mathrm{SO} (14) \times E_8$                                                        & $ 15\widetilde{\mathcal{F}}^T_{0,\alpha} (\tau_2 , N) +  \widetilde{\mathcal{F}}^T_{0,\beta} (\tau_2 , N)$                                                      \\ \hline
\end{tabular}
}
\caption{List of the allowed theories,  gauge groups\protect\footnotemark\ and the tachyonic terms}
\label{table1}
\end{table}

\section{Discussion}

The results obtained in this article are physically intriguing.   Starting from the two consistent heterotic string theories,  we have shown that the gauge symmetry can be broken by a $\mathbb{Z}_N$ orbifold in eight different ways for generic values of $N$.   
The results are summarized in Table \ref{table1}. 

\footnotetext{In writing down the gauge groups,  we have not considered the possibility of these symmetries being enhanced further since the specific form of the groups is not our main concern here.}

As mentioned in the introduction, the existence of different possibilities for implementing the replica method in heterotic string theory could be an indication of different superselection sectors that are allowed by modular invariance.  Embedding the spin connection into the gauge connection to ensure modular invariance of the world-sheet theory is closely related to satisfying the Bianchi identity $\dd H = R \wedge R - F\wedge F$ for the three-form Kalb-Ramond field strength $H$ in the low-energy spacetime theory. It would be interesting to examine the low-energy implications of our results for the possible superselection sectors in the  path integral for the gauge fields following the analysis in  \cite{Buividovich:2008gq,  Donnelly:2011hn,  Casini:2013rba,  Ghosh:2015iwa,  Aoki:2015bsa,  Donnelly:2015hxa,  Blommaert:2018oue, Moitra:2018lxn,  Moitra:2022ooc,  Ball:2024hqe} taking into account the Bianchi identity.

From a technical point of view,  it is remarkable that all the different theories we have examined admit the desired analytic continuation of the tachyons.  There were left-moving tachyonic states with the energy $(-1 + k/N)$.   If such states were present in the physical spectrum, then we would have found tachyonic behavior even in the physical regime. Such states always turned out to be projected out by level-matching or $\mathbb{Z}_N$-invariance as is also known to happen for the ten-dimensional string for the NS sector.  In our previous investigations,  we had found that for the tachyon spectrum, level-matching and $\mathbb{Z}_N$-invariance were concomitant. That is no longer the case here. Another interesting common feature for all the theories is that the net tachyonic contribution can be written in a form
\begin{equation}
\widetilde{\mathcal{F}}^T_0 (\tau_2, N) = (2^n -1) \widetilde{\cal F}^T_{0,\alpha} (\tau_2, N) + \widetilde{\cal F}^T_{0,\beta} (\tau_2, N),
\end{equation}
for some positive integer $n$.   The details of the contributions to the tachyonic spectrum vary across cases,  but the final result is always of this form.

There are many ways to extend the results of this paper.  For instance,  one could also consider compactifications of  heterotic string theories to lower dimensions \cite{Narain:1985jj,  Narain:1986am}.  The previous work \cite{Dabholkar:2023yqc} suggests that the resulting theories would be tachyon-free under an analytic continuation.  The landscape of possibilities in the exploration of quantum entanglement in string theory would certainly be very rich.

\bibliographystyle{JHEP}
\bibliography{heterotic}

\end{document}